\documentclass[aps,pra,reprint,noshowpacs,superscriptaddress]{revtex4-1}   
\usepackage[T1]{fontenc}
\usepackage[latin9]{inputenc}
\usepackage{geometry}	
\usepackage{mdframed}	
\usepackage{framed}	
\usepackage{graphicx}
\usepackage[above,below]{placeins}	
\usepackage{times}
\usepackage[table]{xcolor}
\definecolor{shadecolor}{rgb}{.6,.6,.6}
\begin{document}

\title{Guided and Unguided Student Reflections}
\author{Amanda Matheson}
\affiliation{Department of Physics, Colorado School of Mines}
\author{Laura Wood}
\affiliation{Department of Physics, Seattle Pacific University}
\author{Scott Franklin}
\affiliation{School of Physics and Astronomy, RIT, 1258 Carlson Building, 84 Lomb Memorial Drive, Rochester, NY 14623}

\begin{abstract}
Self-reflection is important metacognitive skill, enabling students to
build coherence into their learning and embed content in a broader
context. While various pedagogical techniques exist to encourage
student reflection, little research has examined the differences
between formally guided, partially guided and unguided
reflections. This study focuses on student responses to online Guided
Reflection Forms (GRFs) from students in a first-semester non-physics
class and, separately, a sophomore-level Vibrations \& Waves course
for physics majors, and compares these guided reflections with
partially guided and unguided journals from a STEM enrichment summer
program for incoming college students. A previously developed coding
scheme was used on guided reflections and the {\bf LIWC}
computational linguistics tool used to confirm the distinct nature of
the categories. A new coding scheme was created and validated for the
unguided journals. We find that both guided and unguided reflections
elicit metacognitive and reflective practice, although of measurably
different frequencies and kinds.
\end{abstract}

\maketitle

\section{Introduction}

A number of studies (e.g. \cite{Bringle99, Scott07, Westberg01,
  Kori14}) have found metacognitive reflection and journaling to
increase student performance and understanding. These need not be
directly related to content; Mason \cite{Mason15} found that students
who set specific goals about future growth, identified theory and
tools to achieve these goals, and reflected regularly on progress
toward those goals demonstrated greater well-being and large
improvements in overall grades. Journals can encourage students to
reflect on their beliefs, thoughts, and actions, in the process
increasing self-efficacy and important sense of self.

Reflections in a classroom setting can be either {\it guided} by
explicit questions or prompts \cite{Dounas-Frazer15} or
unguided. Unstructured journals run the risk of being used as a simple
log book or checklist, \cite{Boud98, Prinsloo11} or subject to student
beliefs that affective or personal reflections are inappropriate in an
academic setting. Guided reflections may help students increase their
reflective sophistication; Kori \cite{Kori14} found that, after
working through a scaffolded reflection that built upon each step of
an experiment, students moved from simpler descriptions to more
sophisticated justifications and critiques of their experimental
methods and analyses.

There have been few attempts to systematically characterize guided or
unguided journals, and virtually no attempts to compare the
two. Dounas-Frazer and Reinholz \cite{Dounas-Frazer15} developed an
online {\it Guided Reflection Form} that encourages specific
reflections on weekly content. In addition, they developed a rubric
for codifying student statements, finding students forthcoming with
simple narrative statements but struggling to articulate concrete
plans to overcome setbacks. Kori \cite{Kori14} developed a similar
categorization scheme for his work on elementary and high-school
student responses to open-ended, yet prompted questions, although his
work was limited to students' understanding of their reasoning process
and did not include personal or non-academic experiences.

\section{Study Design}
\subsection{Instructional Contexts}
Guided Reflection Forms (GRFs), adapted from \cite{Dounas-Frazer15}
were completed weekly by forty-one first-year STEM majors taking the
introductory level ``Metacognitive Approaches to Scientific Inquiry''
({\it Metacognition}) and thirty-eight physics and engineering majors
enrolled in the sophomore-level ``Vibrations \& Waves'' ({\it V\&W})
course. GRFs prompt students to answer a series of three questions:
\begin{enumerate}
\item Describe an experience from the past week that you would like to
    improve upon in the future,
\item Select strategies that you used to overcome the difficult
  situation, and
\item Describe how you would improve upon the experience in the
  future.
\end{enumerate}

Over the course of the semester, this yielded 399 individual
submissions from the Fall 2015 section of {\it Metacognition} and 271
submissions from the Spring 2016 section of {\it V\&W}. Each
submission received a personal e-mail response from the instructor
that affirmed the student's reflections and offered suggestions and
resources to help the student implement the desired improvements.

The {\it Metacognition} students were all first-generation or
deaf/hard of hearing incoming STEM majors taking part in a program,
Integrating Metacognitive Processes and Research to Ensure Student
Success (IMPRESS), \cite{Franklin17} designed to increase
retention. The course introduces students to a variety of
metacognitive and affective topics including mindset, self-assessment,
stereotype threats, and the impacts of micro- and
macro-aggressions. The course is co-taught by an environmental
scientist and a physicist (the last author); the physicist also taught
{\it V\&W}.  {\it V\&W} is taught in an active learning setting, and
covers traditional content from \cite{French}. 

The IMPRESS program includes a two-week summer program for
deaf/hard-of-hearing (DHH) and first generation incoming STEM
majors. Twenty participants conduct experiments and develop models for
climate change as context for exploring and developing metacognitive
and reflective practices. Journals from IMPRESS 2014 and 2015 were
examined as part of this study. In 2014, journals were unguided, with
students only shown examples of topics they could write about. In
2015, students were shown the following set of questions, adapted from
the Guided Reflection Forms:

\begin{itemize}
\item What do you think you learned about thinking style during this
  exercise?
\item What mental resources did you use during the exercise?
\item Think of a specific mental skill you could improve on that would
  you in this exercise.
\item Describe the kinds of skills you think would help you make the
  above improvements.
\item What strategies did you notice your classmates using?
\item How did you group members' strategies impact your approach to
  the problem?
\end{itemize}
\noindent The 2014
Students completed their journals each evening in a diary-style
notebook (e.g. not a lab notebook). Importantly, they did not have the
prompts in front of them when responding. As a result, students took
different approaches to completing their journals. Some would number
each response and answer in order, while others wrote complete
paragraphs that integrated their responses to all the questions.
Students did not receive feedback on their journals in either year.

For our study, 2014 IMPRESS journals are considered {\it unguided},
2015 journals are {\it partially guided}, and GRFs the most guided of
the three methods examined.

\section{Guided Reflections}

GRF submissions were randomly scrambled to prevent the possibility of
bias, with all name, date and course data removed. Two researchers
separately coded each statement by the coding rubric developed by
\cite{Dounas-Frazer15} (see Table~\ref{GRFcode}). Together, these
categories accounted for 40\% of all student statements in both {\it
  Metacognition} and {\it V\&W}.

\begin{table}
\caption{GRF analysis codes developed in \cite{Dounas-Frazer15}}
\label{GRFcode}
\begin{tabular}{| l | l |}
\hline 
{\bf Coding Category} & {\bf Description} \\ \hline
{\it Narrative} & literal descriptions of events \\ \hline
{\it Growth} & broad goals for improvement \\ \hline
{\it Action} & specific goals/steps for improvement \\ \hline
{\it Achievement} & specific desired achievements \\ \hline
\end{tabular}
\end{table}

GRF codes were entered into NVivo, and an Inter-Rater Reliability test
conducted. Given the low coding density per response, a Kappa
coefficient within the range of 0.4-0.7 was considered satisfactory,
with $\kappa >0.7$ representing nearly perfect agreement. Any
categories with $\kappa <0.6$ were discussed until an agreement was
reached. When all the GRFs were considered together, $\langle \kappa \rangle = 0.6$, an acceptable
value for data with low coding density.\cite{kappa}

\subsection{Analysis: coding}

The first comparison is between the guided reflections of incoming
STEM majors in the Metacognition course with the second-year physics
majors in Vibrations \& Waves. A t-test of the frequency of coded
statements by week and across the entire semester revealed no
statistically significant differences between the two sets of
statements. 

\begin{figure}
\begin{center}
\includegraphics[width=3.5in]{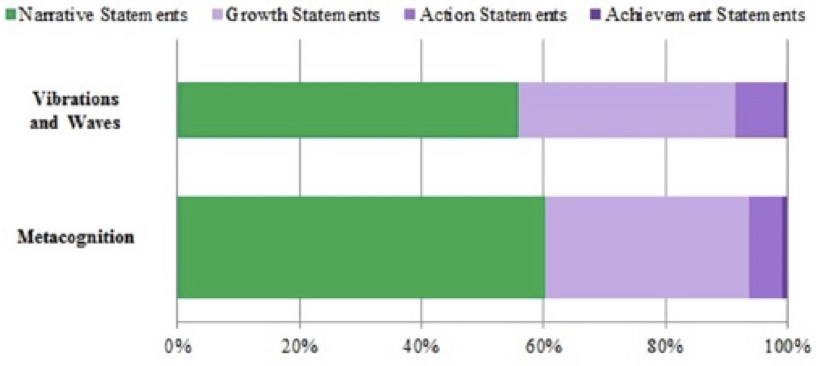}
\end{center}
\caption{Frequency of sentences of different types, normalized by
  total number of coded sentences. The introductory level {\it
    Metacognition} course and the sophomore-level {\it Vibrations \&
    Waves} physics course show nearly identical prevalence of
  narrative statements and scarcity of Action statements.}
\label{fig1}
\end{figure}

The distribution of coded statement frequency in the two groups is
shown in Fig.~\ref{fig1}. Both groups of student reflections are
dominated by narrative statements, which make up almost 60\% of all
coded statements. This is followed by broad growth statements, which
comprise almost 30\% of all statements. Very few of the statements
($<20\%$) are either specific action or achievement statements. This
supports the more general conclusion that students prefer to
articulate less specific goals and narratives and struggle to generate
concrete plans to address obstacles that arise.

There is evidence that GRF prompts elicit the desired types of
responses.  87\% of responses to the prompts "Is there anything else
that you would like to share," and "Describe a specific experience
from last week that you would like to improve upon" were Narrative
statements. Similarly, 81\% of responses to the prompts "Comment on
your experience using these resources last week" and "Describe an
aspect of this experience that you can improve in the future" were
goal statements. Action and Achievement statements are seen in roughly
equal proportion in response to goal-inducing questions. These results
are comparable to those seen in \cite{Dounas-Frazer15}.

\subsection{Analysis: Linguistic Inquiry and Word Count (LIWC)}

In recent years, computerized linguistic analysis, in which
collections of text are compared with established lexicons, has become
more accessible. Tausczik and Pennbaker \cite{Tausczik2010} have
created an online tool, the Linguistic Inquiry and Word Count (LIWC)
\cite{LIWC} that automates the procedure. LIWC draws on dominant
theories in psychology, business and medicine to correlate language
with psychological state. It thus allows the researcher to
characterize text by emotion, thinking style and social concerns.

We analyzed all student reflections with LIWC to see if the coding
rubric could be characterized with distinct LIWC ``fingerprints.''
Three LIWC categories --- Analytical Thinking, Authenticity and
Emotional Tone --- showed statistically significant differences across
categories. Analytical thinking reflects the degree of formality,
logic and hierarchical thinking. Authenticity evaluates the degree of
honesty, personal investment and disclosure. Emotional tone is
correlated with style: positive, upbeat language is characterized as a
high emotional tone, while anxiety, sadness or hostility receives a
lower score. (A fourth category, Clout, reflecting confidence and
expertise was not present in our data.) Each category is rated on a
scale from 0-100, with 50 representing neutrality.

\begin{table}
\caption{LIWC category scores for statements coded as Narrative,
  Growth, Action and Achievement. Shaded boxes indicate scores
  significantly different from the overall mean. Each code has a
  distinct combination of LIWC category scores, validating the coding
  as unique and differentiable.}
\label{table1}
\begin{tabular}{| l | c | c | c|}
\hline 
{\bf Coding Category}& {\bf Analytic}& {\bf Authenticity}& {\bf Tone}
\\ \hline
{\bf Narrative} & \cellcolor{gray!25}53 & 96 & \cellcolor{gray!25}24 \\ \hline
{\bf Growth} & 68 & 90 & 71 \\ \hline
{\bf Action} & 69 & 99 & \cellcolor{gray!25}43 \\ \hline
{\bf Achievement} & 72 & \cellcolor{gray!25}67 & 71 \\ \hline
\end{tabular}
\end{table}

Table~\ref{table1} shows the LIWC coding score for the collective
statements in the four categories, and suggests the following
classifications. Narrative statements are marked by a slightly lower
Analytic score, as they are more personal recollections, and a
distinctly lower (more negative) Tone. This is explained through the
GRF prompt, which asks students to reflect on events or situations
that they would like to improve. While the Narrative fingerprint has
an obvious interpretation, the others are less intuitive. Action
statements are characterized by lower Tone, although not as low as the
Narratives, which suggests that students either are unsure of how to
proceed or are pessimistic at their chances of successfully taking
these actions. Achievement statements, which put forth specific and
desired achievements, have similar reason and tone as others, but
lower authenticity scores. This is quite surprising, and suggests
students struggle to articulate genuine goals. Finally, Growth
statements, which are broader, less specific goals, are characterized
by generally positive scores on all sub-categories.

\section{Partially Guided and Unguided journals}

Shifting attention from the formal classroom, we now look at student
journals from the two-week pre-matriculation summer experience. Twenty
journals from each of two year's (2014 \& 2015) experiences were
collected. Journal entries from the summer focused on metacognitive
aspects of student experiences and reflections, with less attention
paid to overcoming setbacks and long-term goals. The GRF coding rubric
was modified to account for the different reflections seen in the
journals.  As part of this process, two researchers coded journal
entries using the existing GRF scheme and then discussed the
suitability and made agreed-upon changes.

The characterization scheme was also inspired by \cite{Kori14}, whose
scheme forms a loose hierarchy of metacognitive sophistication. Codes
are shown in Table~\ref{jcode}, ordered in increasing
sophistication.  Unguided and partially guided journals resulted in a
much wider range of topics than the formal GRFs, and may include
discussions of other people's approaches or the program activity. To
account for these different statements, the new coding scheme is
considerably broader than that applied to GRFs.

\begin{table}
\caption{Coding rubric for unguided and semi-guided journals, in order
  of nominally increasing metacognitive sophistication.}
\begin{tabular}{| l | p{2.25in} |}
\hline {\bf Coding Category} & {\bf Description}
\\ \hline\label{jcode} {\it Description} & narration of an activity
accompanied by little to no opinions or evaluations \\ \hline {\it
  Logic/rationale} & explicit or implied statements of why a strategy
or approach was chosen \\ \hline
{\it Evaluation} & opinions of an activity or
choice of strategy, accompanied with backing reasons \\ \hline
{\it
  Discussion} & describing alternative solutions \\ \hline
{\it Patterns of thought} & metacognitive descriptions of mental
patterns or strengths or weaknesses \\ \hline
\end{tabular}
\end{table}

The journals were coded in the same manner as the GRFs--- each
researcher coding two thirds --- with an average Kappa coefficient of
0.82 for the 2014 journals and 0.63 for the 2015 journals.

\subsection{Analysis}
For the 2014 journals, students were given no explicit instructions in
what to write. In 2015, students were given prompts adapted from the
GRFs. Perhaps not surprising, the 2015 journals were shorter and more
focused than the 2014 journals. While many students wrote lengthy
reflections including many anecdotal stories of life outside of the
IMPRESS program in 2014, the same was not true in 2015. In addition to
being shorter, the 2015 journals contained a statistically significant
smaller proportion coded statements.

The fraction of Narrative statements in the 2015 journals is also
lower than that in the 2014 journals (47\% vs. 65\%). Descriptive
statements are objective recollections of events, and are thus
analogous with GRF Narrative statements. Recalling that 58\% of
statements in the formal GRFs were Narrative, it is surprising that
that the journal prompts produce a result significantly different from
the GRFs. The 2015 journals also contains more Evaluation, Patterns of
Thought and Discussion statements, and fewer Logic statements.

There is evidence that the student journals from both sets are
considerably richer than the classroom GRFs. More than 80\% of
journals (in both sets) contained statements that were coded in four
of the five five categories, and 60\% had statements that encompassed
all five categories. Conversely, only 50\% of GRFs had statements in
at least three of the four categories, and only 4\% had statements in
all four. The explicit prompts in the Guided Reflection Forms, perhaps
combined with the explicit classroom framing, resulted in reflections
that were more constrained and less personal.

\section{Conclusions}

In this paper, we have examined different methods at encouraging
student reflections. We find that online Guided Reflection Forms are
successful at eliciting reflections, with little difference seen in
courses of different level or student population. Prompts succeed in
helping students focus, stay on topic, and engage in metacognitive
reflection.

Online GRFs present are an efficient mechanism to prompt, collect and
respond to student reflections. Nevertheless, large-scale analysis and
synthesis remains time-consuming. We apply a new tool to this task,
Linguistic Inquiry and Word Count (LIWC) discourse analysis, and find
that the previously developed and validated codes can be uniquely
characterized. Different codes are captured by distinctly different
affective words, varying in degree of formality, personal investment,
and tone. LIWC makes it possible to now study a variety of new,
interesting questions, such as changes in student reflection and
emotional state (as reflected in GRFs) over the course of a semester,
year or academic career.

Journals outside the formal classroom environment, particularly when
unguided, produce a different type of reflection, one considerably
more diverse in types of statements. Unguided journals are longer,
with students exploring a wider range of topics and, surprisingly,
giving more narration. In both formal and informal environments,
prompts are effective at focusing student reflections to desired
topics, and open the possibility for online scaffolding exercises that
lead to greater metacognitive awareness and associated curricular
gains.

Future work could proceed along two different directions. A key
feature of GRFs is timely and personalized instructor feedback. This
feedback is also a rich data set that can be analyzed for linguistic
characteristics. Correlating the nature of instructor feedback with
subsequent student submissions can provide guidance for instructors
looking to improve student metacognition. Separately, student
reflections can be compared with classroom performance to look
directly for a link between metacognitive sophistication and content
learning.

\acknowledgments{We gratefully acknowledge many useful discussions
  with Eleanor Sayre and members of PEER-Rochester, part of the
  Professional development for Education Researchers series. Dimitri
  Dounas-Frazer and Daniel Reinholz first brought GRFs to our
  attention, and provided support and advice as this project
  developed. We are grateful for the support we received from student
  and faculty participants in RIT's REU in DBER Modeling who provided
  guidance and support. Finally, we acknowledge support from the
  National Science Foundation DUE \#1359262 and DUE \#1317450.}

\end{document}